# Proximity-induced high-temperature superconductivity in topological insulators $Bi_2Se_3$ and $Bi_2Te_3$

Parisa Zareapour[1*], Alex Hayat[1,2*], Shu Yang F. Zhao[1], Michael Kreshchuk[1], Achint Jain[1], Daniel C. Kwok[3], Nara Lee[3], Sang-Wook Cheong[3], Zhijun Xu[4], Alina Yang[4], G. D. Gu[4], Shuang Jia[5], Robert J. Cava[5] and Kenneth S. Burch[1]

[1]Department of Physics and Institute for Optical Sciences, University of Toronto, 60 St. George Street, Toronto ON, M5S 1A7, Canada

[2]Centre for Quantum Information and Quantum Control, University of Toronto, 60 St. George Street, Toronto ON, M5S 1A7, Canada

[3]Rutgers Centre for Emergent Materials and Department of Physics and Astronomy, Rutgers University, 136 Frelinghuysen Road, Piscataway, New Jersey 08854, USA

[4]CMP&MS Department, Brookhaven National Laboratory, Upton, New York 11973, USA

[5]Department of Chemistry, Princeton University, Princeton, NJ 08544, USA.

---

[*] These authors contributed equally to this work.

**Abstract**

Interest in the superconducting proximity effect has been reinvigorated recently by novel optoelectronic applications as well as by the possible emergence of the elusive Majorana fermion at the interface between topological insulators and superconductors.

Here we produce high-temperature superconductivity in $Bi_2Se_3$ and $Bi_2Te_3$ via proximity to $Bi_2Sr_2CaCu_2O_{8+\delta}$, in order to access increasing temperature and energy scales for this phenomenon. This was achieved by a new mechanical bonding technique we developed, enabling the fabrication of high-quality junctions between materials, unobtainable by conventional approaches. We observe proximity-induced superconductivity in $Bi_2Se_3$ and $Bi_2Te_3$ persisting up to at least 80K – a temperature an order of magnitude higher than any previous observations. Moreover, the induced superconducting gap in our devices reaches values of 10mV, significantly enhancing the relevant energy scales. Our results open new directions for fundamental studies in condensed matter physics and enable a wide range of applications in spintronics and quantum computing.

**Introduction**

Superconductivity can be produced locally in various materials by placing them in direct contact with a superconductor[1], resulting in unique combinations of distinct properties that do not coexist otherwise[2]. This proximity effect has been proposed as a powerful resource for practical applications in optoelectronics[3], and predicted to produce the elusive Majorana fermion[4] by combining a superconductor with a topological insulator (TI)[5,6,7,8,9,10]. These goals have been pursued actively including the recent demonstrations of novel semiconductor light sources[11] and induced superconductivity in TIs[12,13,14,15,16] such as $Bi_2Se_3$ and $Bi_2Te_3$, which have been shown to have topologically protected surface states[17,18,19]. Nonetheless, all such experiments performed to date have employed low critical temperature (low-$T_c$) materials that require extreme cooling, and the important ratio of the superconducting gap to the Fermi energy in such systems is very small. Inducing superconductivity with the larger energy scales and higher temperatures possible in high-$T_c$ superconductors[20] can enable future realizations of more practical devices as well as greatly expand the ability to study the underlying physical phenomena[21].

A widely used approach for the demonstration of proximity-induced superconductivity in larger-scale devices is based on conductance spectra at superconducting-normal (S-N) interfaces[22]. The proximity-induced superconductivity was established in low-$T_c$ junctions with a metal[23,24], or a semiconductor[25,26] as the N contact. Specifically, Andreev reflection features with the width corresponding to the induced gap in the normal material, along with a reduced gap in the superconductor, were shown to manifest the signature of the proximity effect[23,24,27,28]. Low-$T_c$ superconductivity has been observed recently in $Cu_xBi_2Se_3$[29,30,31], and a corresponding theory of unconventional superconductivity has been developed[32,33]. These low-$T_c$ studies provide the evidence for existence of interactions in TIs, enabling the possibility of inducing a superconducting gap in TIs in proximity with high-$T_c$ materials as well.

Here we generate proximity-induced superconductivity in $Bi_2Se_3$ and $Bi_2Te_3$ by combining them with a high-$T_c$ superconductor $Bi_2Sr_2CaCu_2O_{8+\delta}$ (Bi-2212), and characterize the induced superconductivity by studying the resulting current and differential conductance across the interface. Andreev reflection is observed as an excess current and an increase in differential conductance when cooling below $T_c$. The widths of the central features in the Andreev spectra do not reflect the full gap of Bi-2212, but rather the induced gap in the $Bi_2Se_3$ and $Bi_2Te_3$ due to the proximity effect. Furthermore we observe additional features in the Andreev spectrum due to the reduced gap in the Bi-2212 at the interface - consistent with previous studies of the proximity effect with low-$T_c$ superconductors[23,24,27,28]. This is a natural consequence of the extension of the superconducting order parameter into the $Bi_2Se_3$ and $Bi_2Te_3$, and confirms the existence of the proximity effect in our junctions. The Andreev scattering spectrum we observe shows good agreement with a recently developed d-wave theory[34], modified to account for the high-$T_c$ order parameter profile in the proximity region.

## Results

**Sample characteristics:** High-quality bulk $Bi_2Se_3$, $Bi_2Te_3$ and optimally doped Bi-2212 crystals were cleaved, resulting in atomically flat surfaces over large areas (Fig. 1 a, b), and were subsequently mechanically-bonded (Fig. 1 c, d, e, f) in a dry atmosphere. These low-resistance tunnel junctions where then probed by conductance spectroscopy measurements including current and differential conductance versus voltage (Fig. 1 g). The conductance spectroscopy experiments were performed using four-point probe measurements in a liquid He flow cryostat at different temperatures ranging from 295K to 4.5K. Our mechanically-bonded tunnel junction method was also verified to perform successfully on a Bi-2212/graphite junction (Supplementary Figure S1), resulting in typical S-N tunnelling spectra seen previously for Bi-2212 with point contact or scanning tunnelling microscopy[35].

**Bi-2212/Bi$_2$Se$_3$ experiments:** The first set of experiments was performed on Bi-2212/Bi$_2$Se$_3$ junctions. DC current along the c-axis versus voltage (I-V) characteristics (Fig 2 a) reveal excess current below the Bi-2212 T$_c$ (~85K) due to the Andreev reflection at the S-N interfaces[36], consistent with the model developed for anisotropic superconductors[34]. The observed Andreev reflection indicates that we have achieved a surprisingly low-barrier between Bi-2212 and Bi$_2$Se$_3$. Furthermore, this is the key mechanism for the superconducting proximity effect[37], suggesting the existence of a proximity-induced superconducting region at the Bi-2212/Bi$_2$Se$_3$ interface. Just below T$_c$, the excess current, I$_e$, reaches the maximal value similar to the normal-normal (N-N) interface current (measured for T > T$_c$), resulting in total current nearly twice that of the N-N interface due to the contribution of the Cooper pairs when Andreev reflection occurs. A different measurement - AC differential conductance below T$_c$, $(dI/dV)_S$, divided by the normal state conductance $(dI/dV)_N$ measured at 105K, confirms the DC I-V measurement. Specifically, the AC differential conductance reveals a zero-bias conductance peak due to Andreev reflection below T$_c$ (Fig. 2 b). At lower temperatures (around 60K), reduction of higher-bias current was observed, and at the same temperatures additional higher-bias features appeared in the differential conductance measurement. These features result from the gap reduction in Bi-2212 and the induced gap in Bi$_2$Se$_3$. However, the detailed structure of these spectral features could be studied only at lower temperatures as discussed in detail below.

To demonstrate the spectral features more clearly, we performed both DC and AC measurements at temperatures well below $T_c$. DC I-V characteristics of the junction (Fig. 2 c) show excess current at lower voltages up to about 13mV with a correspondingly smaller value $I_S \sim 2.7\mu A$. However, two additional step-like features appear near 27mV and 45mV. These features are confirmed by an AC differential conductance measurement (Fig. 2 d), showing a wide conductance feature between -13mV and +13mV, as well as peaks at ±27mV and ±45mV, in good quantitative agreement with the DC measurement (Fig. 2 c inset). Similar results were obtained with Bi-2212/$Bi_2Te_3$ junctions - described below. This conductance spectrum is a clear signature of a proximity-induced superconducting region in the $Bi_2Se_3$.

When the proximity effect occurs (Fig. 3 a (I)), a superconducting gap is induced in the normal material, $\Delta_i$, and near the interface the gap in the superconducting material is reduced from the intrinsic value, $\Delta_0$ (Fig. 3 a (II)), to a smaller one, $\Delta_r$[1,23,24,27,28]. Generally, the superconducting gap is position-dependent along the axis normal to the interface plane, $\Delta(x)$, and Andreev scattering probability is finite in the whole proximity region. However, the scattering can be divided into two main energy ranges: electrons (holes) with energy $|E|<\Delta_i$ will be mainly Andreev reflected inside the proximity-induced superconducting region of the normal material, whereas electrons (holes) with $\Delta_i<|E|<\Delta_r$ will continue to the S-N interface and Andreev-reflect mainly near the material interface. Thus, conductance spectrum taken on an interface exhibiting the proximity effect will show Andreev scattering features corresponding to both gaps[23,24,27,28]. In our experiments, therefore, the central conductance peak is a manifestation of Andreev reflection from the normal region to the proximity-induced superconducting region of $Bi_2Se_3$ with a gap $\Delta_i \sim 13$mV. The reduced gap in Bi-2212 due to the proximity effect appears as a conductance peak at $\Delta_r \sim 27$mV. The change in the zero-bias conductance with temperature is an additional manifestation of the Andreev scattering. Indeed, just below $T_c$ the zero-bias conductance increases to nearly twice its value above $T_c$ (Fig. 3 b). This temperature dependence as well as the spectral shape of the conductance at various temperatures (Fig. 3 b inset) completely rules out any possible heating-related effects, whereas a junction with an increased barrier exhibits no proximity (Fig. 3 c). Andreev reflection between Bi-2212 and normal $Bi_2Se_3$ (or $Bi_2Te_3$) with no proximity would have appeared as a single Andreev feature with a width corresponding to the full gap of Bi-2212, which is completely different from our observations. The full gap appears in our experiment as additional peaks at $\Delta_0 \sim 45$mV, whose magnitude and temperature dependence are quantitatively consistent with previous tunnelling studies of Bi-2212[35] and our high-barrier junctions (Fig. 3 c).

The zero-bias conductance feature has a number of characteristics consistent with Andreev reflection in a proximity-induced region. The height of the $Bi_2Se_3$ Andreev zero-bias conductance feature is nearly twice the normal conductance value due to the Cooper pair contribution (Fig. 2 d and Fig. 3 b), whereas the width is determined by the induced superconducting gap, $2\Delta_i$. The width of the Andreev feature is also manifested in the excess current up to a negative bias $\Delta_i$ in the DC I-V characteristic (Fig. 2 c) - nearly twice as high as the current in the normal state (Fig. 2 a). The probability of Andreev scattering at the interface with the Bi-2212 is reduced by the scattering in $Bi_2Se_3$, resulting in slightly smaller contribution to the conductance at $\Delta_r$. At low temperatures both the induced gap and the reduced gap features in differential conductance result in a step-like I-V curve (Fig. 2 c). With increasing temperature the features in the differential conductance merge into one wide central peak (Fig. 2 b), and the step-like structure in I-V disappears (Fig. 2 a). Therefore at certain higher temperatures the total excess current can be larger than at lower temperatures.

**Theoretical modelling:** For the quantitative theoretical modelling of the effect, we calculated the c-axis Bi-2212/Bi$_2$Se$_3$ and Bi-2212/Bi$_2$Te$_3$ conductance spectra using the S-N transport formalism developed for anisotropic superconductors[34]. The differential conductance below T$_c$ $(dI/dV)_S$, divided by the normal state conductance $(dI/dV)_N$ is given by the half-sphere integration over solid angle[38] $\Omega$ :

$$\sigma(E) = \frac{\int d\Omega \sigma_N \cos\theta_N \sigma_R(E)}{\int d\Omega \sigma_N \cos\theta_N} \quad (1)$$

where $E$ is the quasiparticle energy and $\theta_N$ is the incidence angle (relative to the interface normal) in the normal material, $\sigma_N$ is the conductance from normal to normal material with the same geometry, and

$$\sigma_R(E) = \frac{1 + \sigma_N |\kappa_+|^2 + (\sigma_N - 1)|\kappa_- \kappa_+|^2}{\left|1 + (\sigma_N - 1)|\kappa_- \kappa_+|\exp(i\varphi_- - i\varphi_+)\right|^2} \quad (2)$$

where $\kappa_\pm = \left[E - \sqrt{E^2 - |\Delta_\pm|^2}\right]/|\Delta_\pm|$ and $\Delta_\pm = |\Delta_\pm|\exp(i\varphi_\pm)$ - electron-like and hole-like quasiparticle effective pair potentials with the corresponding phases $i\varphi_\pm$.

In the case of c-axis tunnelling, the hole-like and the electron-like quasiparticles transmitted into the superconductor experience the same effective pair potentials, which have similar dependence on the azimuthal angle $\alpha$ in the *ab* plane $\Delta_+ = \Delta_- = \Delta_0 \cos(2\alpha)$. The total Andreev reflection spectrum is obtained by calculating the reflection and the transmission in the proximity region, followed by reflection at the interface between the two materials. The calculated spectra in this two-stage scattering model with the modified gaps as fit parameters show good agreement with the experimental conductance measurements (Fig. 3 d).

**Bi-2212/Bi$_2$Te$_3$ experiments:** In order to demonstrate the wide applicability of our mechanical bonding technique, we have constructed similar S-N junctions combining Bi-2212 from a different batch of crystals and Bi$_2$Te$_3$ grown by a different group (Princeton U.) than that producing the Bi$_2$Se$_3$ (Rutgers U.). Conductance measurements of low-barrier Bi-2212/Bi$_2$Te$_3$ junctions also clearly show the proximity-induced gap in Bi$_2$Te$_3$ as well as the reduced gap in Bi-2212 (Fig. 4 a). The central Andreev feature due to the induced gap in the spectrum also appears immediately below T$_c$ similar to Bi$_2$Se$_3$, with a conductance increase of nearly twice the normal value (Fig. 4 b), and a width consistent with Andreev reflection in the proximity-induced region. We have constructed several Bi-2212/Bi$_2$Te$_3$ junctions with various induced gap sizes, and the measured spectra agree well with the calculations (Fig. 4 b inset). One of the Bi-2212/Bi$_2$Te$_3$ junctions exhibited particularly large proximity-induced features in differential conductance (Fig. 4 c), and the corresponding current-voltage DC measurement shows excess current as high as 90μA (Fig. 4 d). The proximity related Andreev features and the excess current were observed repeatedly in several Bi-2212/Bi$_2$Se$_3$ (Fig. 4 e) and Bi-2212/Bi$_2$Te$_3$ (Fig. 4 f) devices.

**Discussion**

To rule out any possible alternative interpretations of our results, we considered various complications caused by the planar geometries of our devices, which are different from those based on point contacts that guarantee ballistic transport. In planar junctions, electron scattering can slightly affect the ballistic nature of the transport. This could have some effect on our measurements; nonetheless, it can only result in the observation of features at voltages that are a little higher than the true electron energies. However, this cannot produce an Andreev feature that is narrower than the true gap of the system. Furthermore, this effect cannot produce additional features in the spectra that we observe which are consistent with the reduced gap at the interface. Moreover, to the extent that this effect is relevant it could result in a measured bulk Bi-2212 gap that is bigger than the known gap size from photoemission and tunneling studies. However we observe a bulk gap in accord with previous measurements and that is fairly insensitive to the details of the contact (see Supplementary Discussion and Supplementary Figure S2).

Previous low-$T_c$ S-N planar-junction conductance experiments[39] showed a narrow and small zero-bias conductance peak (several percent increase) due to pair currents. Later, a more detailed interpretation was suggested with a more complex Andreev reflection process including electron-hole phase conjugation[40]. This phase conjugation model, however, predicts a rapid monotonic decrease of conductance at nonzero bias due to the different phases accumulated by quasiparticles. Therefore, this model is not applicable to the wide spectral features with complex shapes corresponding to the Andreev reflection at the induced gap interface observed in our experiments and in previous low-$T_c$ work[23,24,27,28]. Moreover, our experimental results are also different from the low-$T_c$ results[39,40]. Our measurements show a proximity-induced gap feature with a wide region, appearing immediately below $T_c$ with almost twice the normal conductance (Fig. 3 b and Fig. 4 b). Furthermore, our measured Andreev spectra show several distinct features related to the induced gap in the normal material and the reduced gap in the superconductor (similar to ref. 28). At lower temperatures, the central feature in our measured spectra is almost independent of the bias, which is consistent with the proximity-induced gap, and different from the phase-conjugation model[40].

Moreover, any possible *ab*-plane tunneling in our devices can be ruled out for a number of experimental reasons. First, the areal coverage is much greater for c-axis tunneling then for *ab* plane tunneling. In fact, when the topological insulator surfaces were rougher we found no proximity effects, confirming that additional area for any possible ab-plane tunneling did not change the spectra significantly. Secondly, in bulk Bi-2122, the c-axis resistivity is very high only in the normal state, while in the superconducting state, all directions reveal zero-resistance. Therefore since the area perpendicular to the c-axis is much larger than the exposed area normal to the ab-plane, the overall current is along the c-axis, eliminating any possible *ab*-plane effects. Lastly, our c-axis tunneling conductance measurements are very different from an *ab*-plane Andreev bound-state peak[41]. Specifically our differential conductance measurements reveal a very wide central Andreev feature, which is almost independent of the bias. This is consistent with Andreev reflection in the normal material from an induced gap and is completely different from an Andreev bound state. Furthermore, an Andreev bound state related peak in conductance should have a height that is independent of the barrier strength – Z. However, in our case, increasing Z results in a very different conductance spectrum shape. This is consistent with elimination of a proximity region at high Z, and is inconsistent with a bound state.

To further verify that the Andreev scattering is due to the proximity induced superconductivity in $Bi_2Se_3$, and dismiss any possible processes occurring in Bi-2212 alone, the Bi-2212/$Bi_2Se_3$ junction barrier was increased mechanically. Rapidly heating the junction to 300K, and cooling it back to 4K in less than 30 minutes achieved this. As expected, the differential conductance measurement performed on the high-resistance junction revealed only spectra typical of a c-axis tunneling measurement on Bi-2212[35], indicating the lack of a proximity effect. Specifically, the superconducting gap of Bi-2212 was observed, however no Andreev peak appeared at the Bi-2212 reduced gap $\Delta_r$, and no conductance features corresponding to proximity-induced gap in the $Bi_2Se_3$ (Fig. 3 c). The differential conductance spectrum for the high-resistance junction showed the superconducting gap of Bi-2212, $\Delta_0$, and its temperature dependence agrees with the values from the low-resistance junction measurement (Fig. 3 b inset) as well as for the graphite junction (Supplementary Figure S1), confirming the origin of the conductance peaks.

Our results present clear evidence of proximity-induced high-$T_c$ superconductivity in $Bi_2Se_3$ and $Bi_2Te_3$. We achieved this via our newly-developed technique of mechanically-bonded junctions, and confirmed it by various experiments. Specifically below $T_c$, we observe Andreev features corresponding to a reduced gap of Bi-2212 appearing concurrently with an induced gap in $Bi_2Se_3$ or $Bi_2Te_3$. The disappearance of these features in the same junctions when the barrier is increased further confirms that the proximity effect is the origin of these features. The proximity-induced superconductivity in $Bi_2Se_3$ and $Bi_2Te_3$ is demonstrated at temperatures at least an order of magnitude higher than previously reported results. The developed mechanical bonding technique may render various future experiments on novel materials, including high-$T_c$ superconductors and TI feasible. Furthermore our proximity demonstration paves the way for practical realization of TI-based devices involving superconductivity, including Majorana fermion based topological quantum computing.

## Methods

**Device Fabrication:** High-quality bulk $Bi_2Se_3$ and $Bi_2Te_3$ crystals were prepared as described elsewhere[42]. Additionally, optimally doped $Bi_2Sr_2CaCu_2O_{8+\delta}$ (Bi-2212) crystals ($T_c \approx$ 85K) were grown by the floating-zone method[43]. Bi-2212/$Bi_2Se_3$ and Bi-2212/$Bi_2Te_3$ junctions were fabricated by using these crystals as the starting materials. The formation of atomically smooth surfaces via cleaving is well established for Bi-2212, $Bi_2Te_3$ and $Bi_2Se_3$. Bi-2212 was cleaved in a dry nitrogen purged box with adhesive tape (Fig 1. c), producing a clean, flat piece of Bi-2212 a few hundred µm thick. However, we have found the $Bi_2Te_3$ and $Bi_2Se_3$ surfaces were not smooth enough when simply cleaved using scotch tape. Thus we cleaved a bulk $Bi_2Se_3$ or $Bi_2Te_3$ crystal in the dry box by sandwiching the crystal between two glass slides with double-sided tape. After applying slight pressure to the glass slides, the top glass slide was lifted off (Fig 1. d). This approach was taken due to the difficulty of producing flat surfaces by cleaving $Bi_2Se_3$ or Bi$_2$Te$_3$ with scotch tape, as was done for Bi-2212. Next, the Bi$_2$Se$_3$ or Bi$_2$Te$_3$ was transferred to a Cu sample holder by attaching a double-sided tape to the Cu and placing one of the glass slides with Bi$_2$Se$_3$ or Bi$_2$Te$_3$ on the sample holder. By lifting off the glass slide, a smooth and fresh surface of Bi$_2$Se$_3$ or Bi$_2$Te$_3$ was left on the sample holder. The Bi-2212 was then attached to the Bi$_2$Se$_3$ or Bi$_2$Te$_3$ by placing the Bi-2212 on the Bi$_2$Se$_3$ or Bi$_2$Te$_3$, applying GE varnish on the corners of Bi-2212 (Fig 1. e). In the fabrication process, no pressure is required and the junction is formed spontaneously. The important point is cleaving the materials in a dry and inert atmosphere (an N$_2$ purged glove box in our case) with a technique that ensures large, atomically smooth surfaces. Contacts were made on the sample using Cu wires and Ag epoxy or Ti/Au (Fig 1. f). This method was inspired by the mechanical exfoliation technique, however no exfoliation is needed here, and bulk cleaved crystals are attached mechanically using GE varnish on the corners. Low-barrier junctions were only revealed when the Bi$_2$Se$_3$ and Bi$_2$Te$_3$ were cleaved using our glass-slide technique. When simply cleaving the samples with scotch tape, all junctions revealed high-barrier behavior, specifically the resistance across the junction was high (> tens of kOhms at room temperature) and no Andreev features were observed upon cooling.

**Measurement Setup Description:** In our setup, four-point probe measurements were performed in a liquid He flow cryostat at different temperatures ranging from 295K to 4.5K. Bias-dependent differential conductance as well as current vs. voltage were measured on the junctions using two lock-in amplifiers (Stanford Research Systems SR810), a DC voltage source (BK Precision 1787B), two DC multi-meters (Hewlett Packard 3457A and Agilent 34401A) and a home-built, shielded AC+DC adder box (Fig 1.g). One of the lock-in amplifiers was used to produce a small AC voltage output at a frequency near 1 kHz. The AC voltage was added to the DC output of a DC power supply using a transformer-based adder. A voltage was applied to the sample, with the resulting voltage measured with a lock-in (the AC part) and a multimeter (the DC part). At the same time, the current was converted to voltage, amplified with a preamplifier (SRS 570) and measured with another lock-in (AC) and a multimeter (DC).

**Acknowledgements**

The work at the University of Toronto was supported by the Natural Sciences and Engineering Research Council of Canada, the Canadian Foundation for Innovation, and the Ontario Ministry for Innovation. The work at Rutgers was supported by National Science Foundation DMR-1104484. The crystal growth at Princeton was supported by the US National Science Foundation, grant number DMR-0819860.

**Author contributions**

P.Z. and A.H. equally contributed to this work. P.Z., A.H. and K. S. B. planned the research. S.Y.F.Z. and A.J. implemented the conductance measurement setup. A.H. and M.K. carried out the theoretical modelling. D. C. K., N. L., S.-W. C., Z. X., A. Y., G. D. G., S.J. and R. J. C. grew the crystals for the devices. P.Z. and A.H. performed the experiments and wrote the paper. K. S. B. conceptualized the experiment and supervised the work.

**Additional information**

**Supplementary Information** accompanies this paper at
http://www.nature.com/naturecommunications

**Competing financial interests:** The authors declare no competing financial interests.

**Figure legends:**

**Figure 1. | Device fabrication and measurement setup. (a)** Atomic force microscope (AFM) image of the crystal surface of $Bi_2Se_3$. The scalebar corresponds to $2\mu m$. **(b)** Larger area AFM scan of the $Bi_2Te_3$ sample, demonstrating the vertical inhomogeneity of the surface is limited to ±2 unit cells. The scalebar corresponds to $3\mu m$. Junction fabrication technique: **(c)** Bi-2212 (BSCCO) crystal is cleaved using scotch tape**. (d)** $Bi_2Se_3$ (or $Bi_2Te_3$) is sandwiched between glass slides with double-sided tapes and the top glass slide is lifted off, cleaving a flat surface. **(e)** The $Bi_2Se_3$ (or $Bi_2Te_3$) is transferred to a Cu sample holder, and the cleaved Bi-2212 crystal is applied to $Bi_2Se_3$ or $Bi_2Te_3$ using GE varnish at the corners. **(f)** Contacts are made with Ag epoxy or evaporated Au/Ti. **(g)** Experimental setup: four-point DC current-voltage and AC differential conductance measurements performed down to 4.5K, using a liquid He flow cryostat with lock-in amplifiers. DC bias from the power supply is combined with the AC signal from the voltage lock-in amplifier in a transformer-based adder.

**Figure 2. | Bi-2212/$Bi_2Se_3$ junction measurements. (a)** DC current-voltage characteristics of the low-resistance Bi-2212/$Bi_2Se_3$ junction for different temperatures above and below $T_c$. Below $T_c$, excess current typical of Andreev reflection at an S-N interface is exhibited. The 70K I-V curve shows the maximal excess current $I_e \sim 4.85\mu A$ – similar to the value of the normal junction (above $T_c$) at the voltage where the superconducting-state current becomes linear (black dashed line). The total excess current is smaller for lower temperatures due to additional gap features. **(b)** AC

Differential conductance $(dI/dV)_S$, normalized by the normal state conductance $(dI/dV)_N$ at 105K, for the low-resistance Bi-2212/Bi$_2$Se$_3$ junction at different temperatures above and below T$_c$. A zero-bias conductance Andreev peak is clearly seen below T$_c$, with additional features in the spectrum appearing at lower temperatures. The curves are shifted for clarity. **(c)** DC current-voltage characteristics of the low-resistance Bi-2212/Bi$_2$Se$_3$ junction for different temperatures well below T$_c$. In addition to the excess current, the low-temperature curves exhibit two distinct steps corresponding to the dips in the differential conductance measured for AC bias. The black arrow shows the induced Bi$_2$Se$_3$ gap, $\Delta_i$, with a current twice that of the normal state at the same voltage (Fig. 2 a), and the red arrow indicates the reduced Bi-2212 gap, $\Delta_r$, while the intrinsic Bi-2212 gap, $\Delta_0$, is shown by purple arrows. The inset shows the DC and the AC differential conductance at 4.5K, indicating the good correspondence between the different measurements of $\Delta_0$ (purple arrow), $\Delta_r$ (red arrow), and $\Delta_i$ (black arrow). **(d)** AC Differential conductance $(dI/dV)_S$, normalized by the normal state conductance $(dI/dV)_N$ at 105K, for a low-resistance Bi-2212/Bi$_2$Se$_3$ junction for different temperatures well below T$_c$. The curves are shifted for clarity. The zero-bias conductance feature is due to the Andreev reflection between the normal and proximity induced superconducting regions in Bi$_2$Se$_3$, where the width of the peak is nearly $2\Delta_i$. The two additional peaks indicate the reduced and the intrinsic Bi-2212 gaps.

**Figure 3 | Proximity induced superconductivity. (a)** A schematic drawing of the junction in two regimes: (I) low-resistance with proximity induced superconductivity in $Bi_2Se_3$ (or $Bi_2Te_3$). Andreev scattering takes place in the whole proximity region with lower energy particles reflected mainly in $Bi_2Se_3$ (or $Bi_2Te_3$), and higher energy ones mainly at the interface with Bi-2212. (II) high-resistance with no proximity induced superconductivity in $Bi_2Se_3$ (or $Bi_2Te_3$) – quasiparticle scattering occurs at the interface. **(b)** Temperature dependence of the zero-bias differential conductance for Bi-2212/$Bi_2Se_3$. Below $T_c$, the Andreev process enhances the conductance by almost a factor of 2. The inset shows the temperature dependence of the reduced Bi-2212 gap $\Delta_r$ blue circles) and the intrinsic Bi-2212 gap $\Delta_0$ in the low-resistance junction (green squares), and in the high-resistance junction (red diamonds). **(c)** Differential conductance $(dI/dV)_S$, normalized by the normal state conductance $(dI/dV)_N$ at 105K, for a high-resistance Bi-2212/$Bi_2Se_3$ junction for different temperatures, demonstrating the typical quasiparticle tunneling differential conductance resulting in a conductance dip in the gapped region. The curves are shifted for clarity. The arrows indicate the intrinsic superconducting gap of Bi-2212, $\pm\Delta_0$, around 45mV. **(d)** Measured (solid red line) and calculated (dashed blue line) 15K differential conductance $(dI/dV)_S$, normalized by the normal state conductance $(dI/dV)_N$ at 105K, for a low-resistance Bi-2212/$Bi_2Se_3$.

**Figure 4. | Bi-2212/$Bi_2Te_3$ junction measurements. (a)** AC Differential conductance $(dI/dV)_S$, normalized by the normal state conductance $(dI/dV)_N$ at 100K,

for a low-resistance Bi-2212/Bi$_2$Te$_3$ junction at different temperatures above and below T$_c$. The curves are shifted for clarity. **(b)** Temperature dependence of the zero-bias differential conductance for Bi-2212/Bi$_2$Te$_3$. Below T$_c$ the Andreev process enhances the conductance by almost a factor of 2. The inset shows 10K measured (solid lines) and calculated (dashed lines) normalized differential conductance for several Bi-2212/Bi$_2$Te$_3$ devices with different values of the induced gap size. **(c)** AC Differential conductance $\left(dI/dV\right)_S$, normalized by the normal state conductance $\left(dI/dV\right)_N$ at 100K, for a large-induced-gap Bi-2212/Bi$_2$Te$_3$ junction at different temperatures. The curves are shifted for clarity. **(d)** DC current-voltage characteristics of the large-induced-gap Bi-2212/Bi$_2$Te$_3$ junction, for different temperatures above and below T$_c$. **(e)** Measured normalized differential conductance for two Bi-2212/Bi$_2$Se$_3$ devices with different values of the induced gap size. The curves are shifted for clarity **(f)** Measured normalized differential conductance for three Bi-2212/Bi$_2$Te$_3$ devices with different values of the induced gap size. The curves are shifted for clarity.

# Figure 1

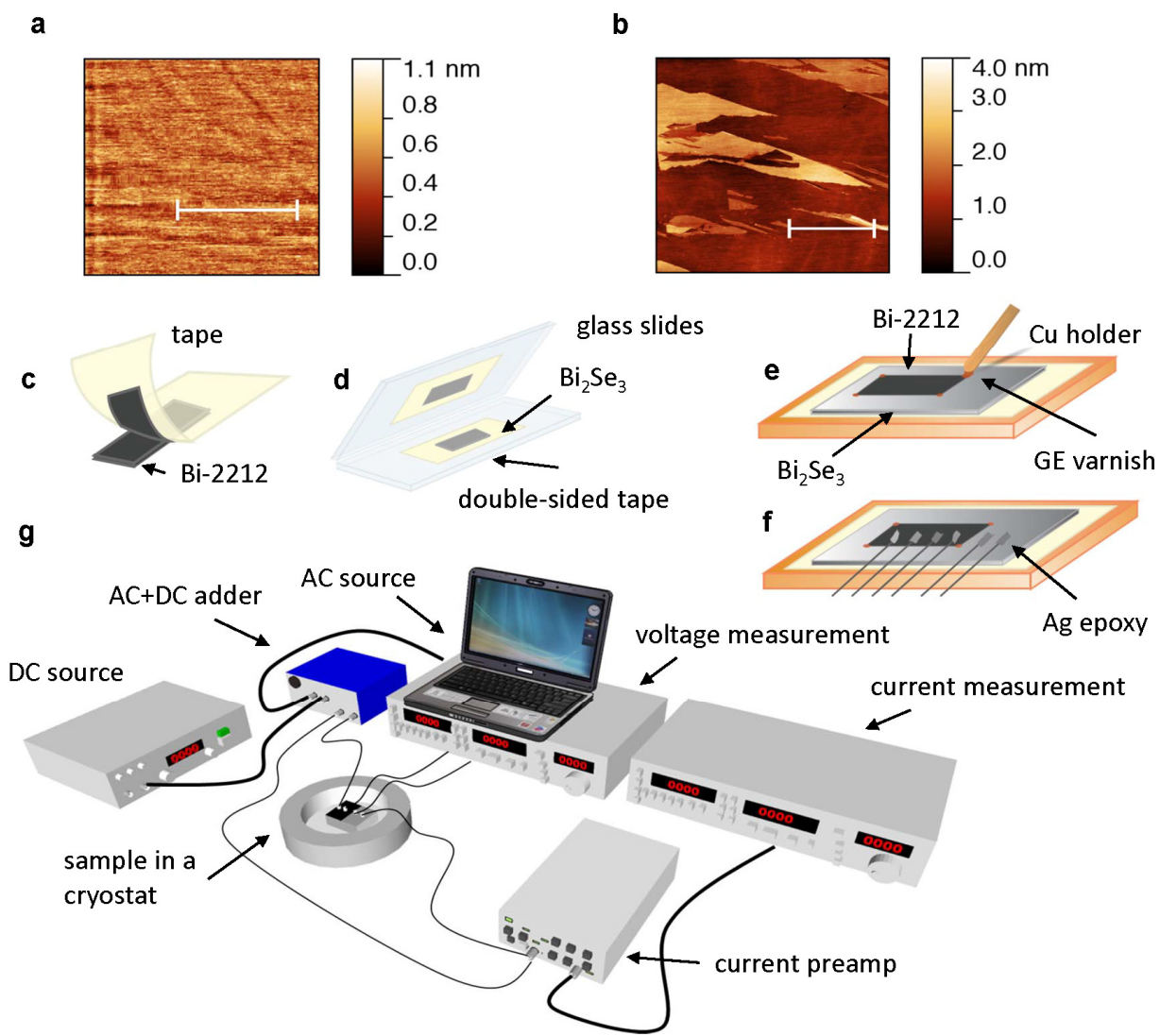

**Figure 2**

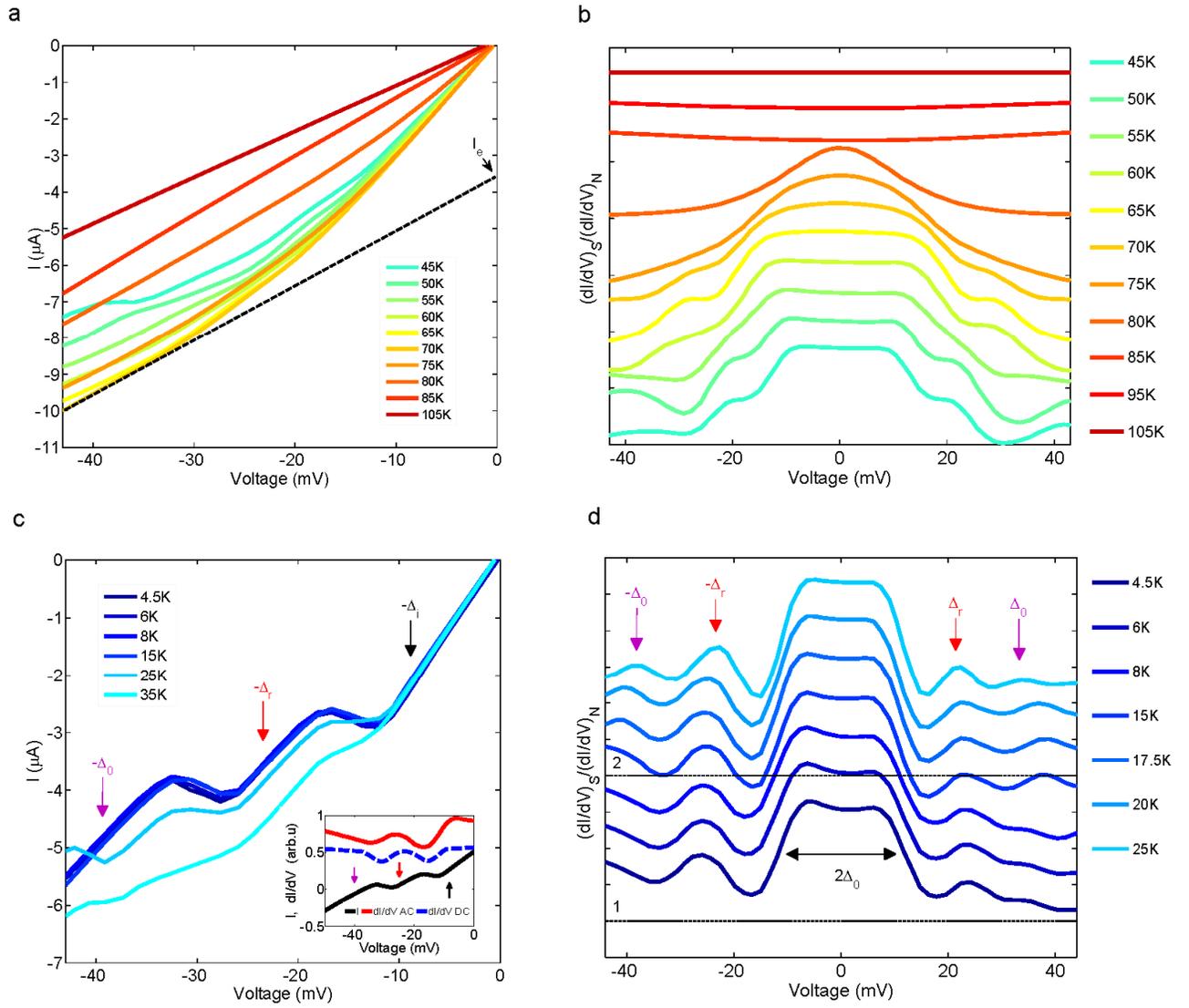

# Figure 3

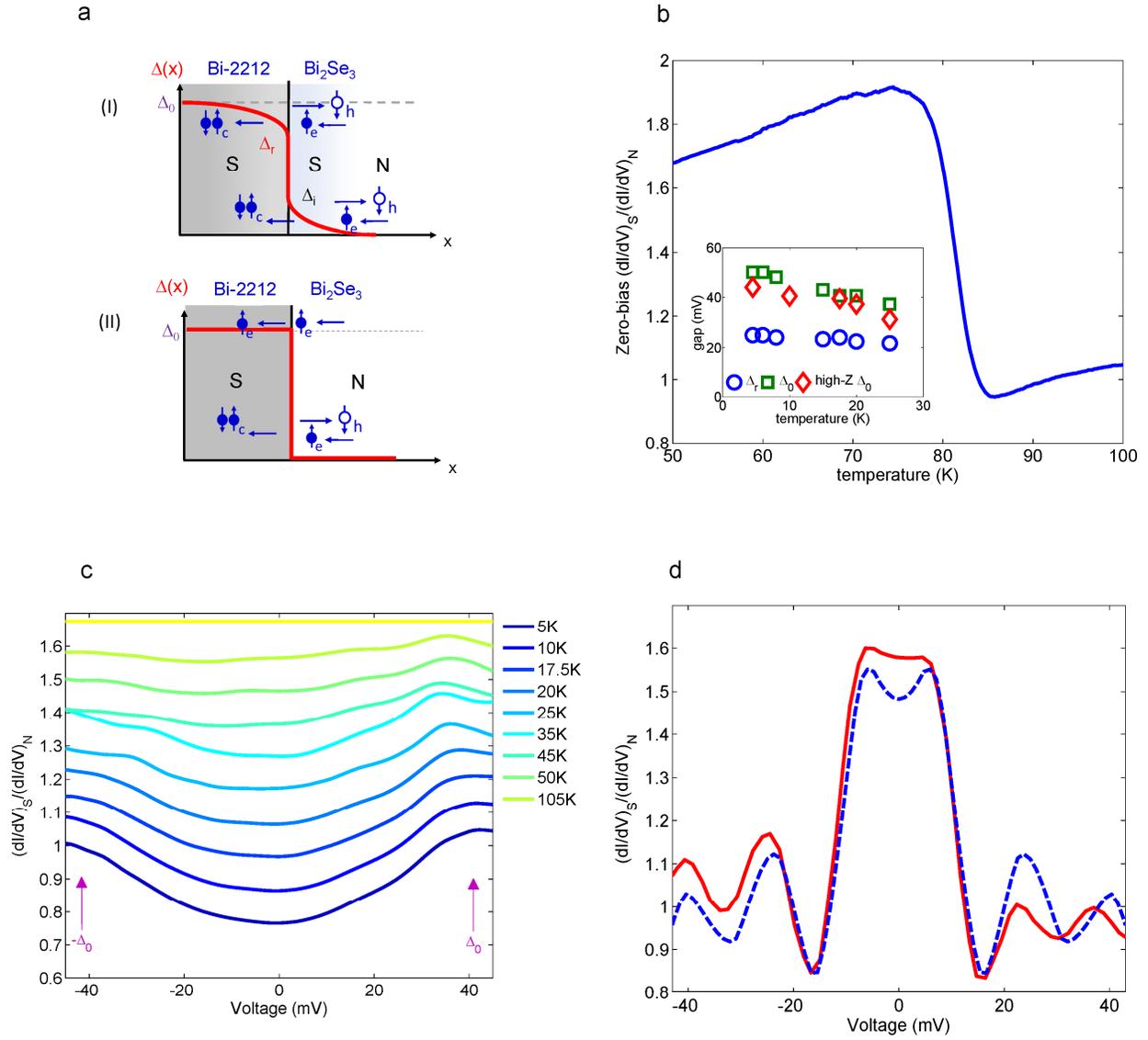

# Figure 4

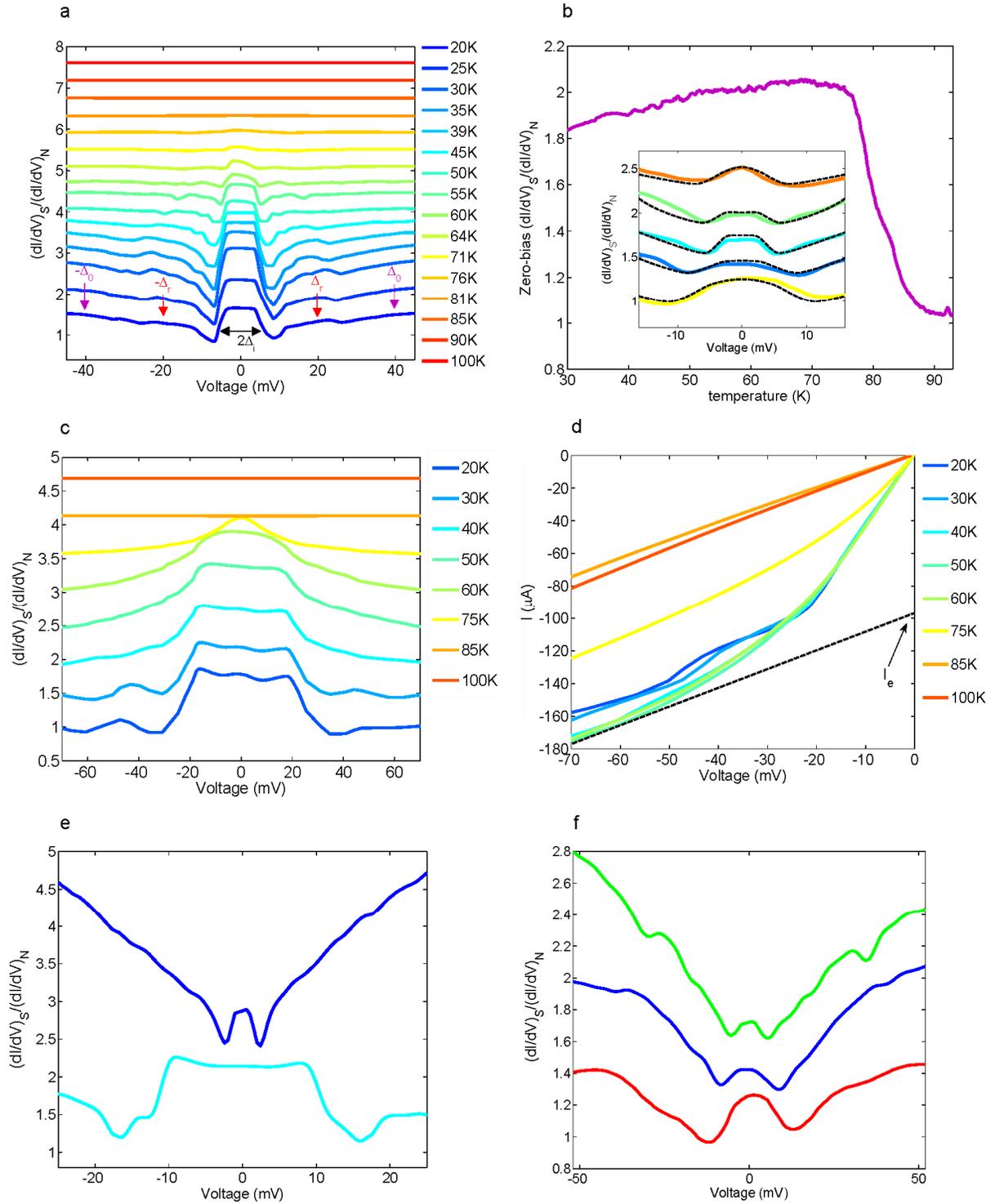

## Supplementary Information

**Supplementary Figures**

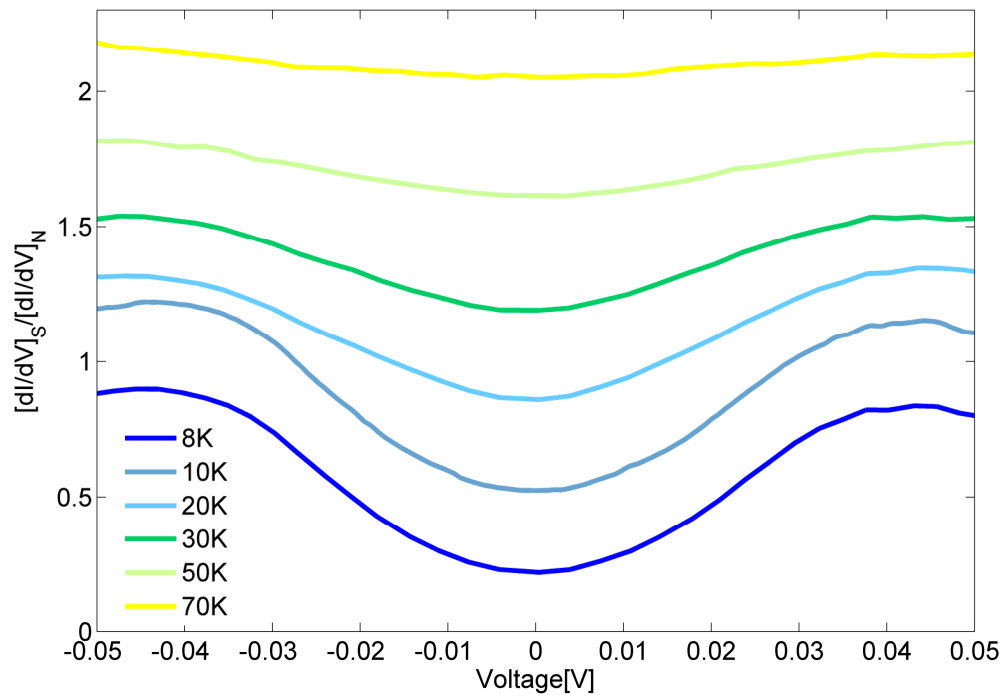

**Supplementary Figure S1 | Bi-2212/graphite measurement.** AC Differential conductance $[dI/dV]_S$, normalized by the normal state conductance $[dI/dV]_N$ at 105K, for a Bi-2212/graphite junction at different temperatures below $T_C$. The curves are shifted for clarity.

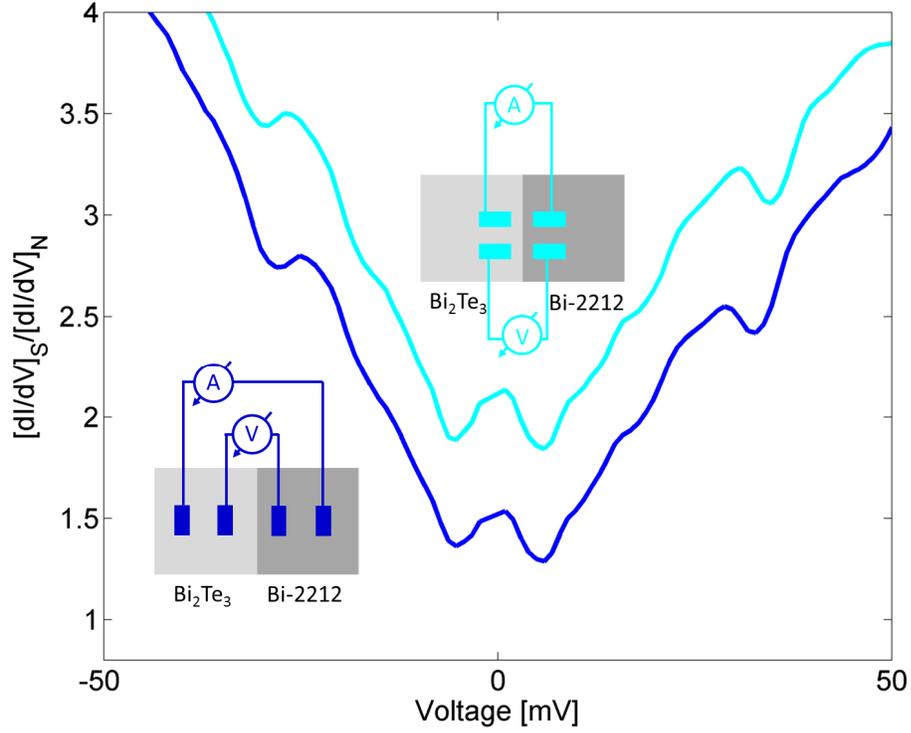

**Supplementary Figure S2. | Contact geometries**. AC Differential conductance $[dI/dV]_S$, normalized by the normal state conductance $[dI/dV]_N$ at 105K, for a Bi-2212/Bi$_2$Te$_3$ junction with two different contact configurations both at 5K. The curves are shifted for clarity. The insets are schematic drawings of the contact configurations.

**Supplementary Discussion**

In order to test the robustness of the macroscopic planar junction based devices to different contact geometries, we constructed a Bi-2212/Bi$_2$Te$_3$ device with two different contact configurations on the same junction (Supplementary Figure S2).

In this experiment, only the contact configuration is varied between the two measurements, and all other conditions of the experiment are similar, which allows effects due to contact configuration to be tested separately. One set of four contacts was fabricated on the same line, while the other set of four contacts was fabricated as two parallel pairs (Supplementary Figure S2 insets). The distances between the contacts are as large as the device itself, allowing any possible geometric effects to be maximized. Our differential conductance measurements show only a small difference at which the high-bias features appear. This difference can be attributed to carrier scattering resulting in slightly non-ballistic transport.

The contact in our junctions is obtained mechanically, and can therefore be somewhat non-uniform. However we have constructed various mechanically bonded devices from Bi$_2$Se$_3$ and Bi$_2$Te$_3$ with various contact geometries and barrier strengths (Fig 2-4, Supplementary Figures S1 and S2). Any possible Josephson networks would have resulted in different features for every junction. Nonetheless all of our devices exhibit similar features in the conductance spectra resulting from the proximity effect at the interface, with trends that evolve with barrier height as predicted by theory.

Bi-2212 can lose some portion of its oxygen content in the fabrication process, and indeed some of the devices we fabricated showed a slightly smaller gap and a slightly reduced $T_c$. However, all of our devices exhibit Andreev reflection spectra with features corresponding to a proximity-induced gap in the normal material for junctions with small resistance. Furthermore, these features always emerge very close to the well-established bulk $T_c$ for our crystals, which has been further confirmed in-situ by performing 4 point resistance measurements on the bulk Bi-2212 in the device. When the resistance of the junction is increased by enlarging the gap between the materials using thermal cycling, the proximity features disappear and a typical S-N tunneling spectrum is observed. Nonetheless, the features observed always appear at the same temperature, even after thermal cycling, indicating that the oxygen content is fairly stable in our devices. All of these indications prove that the Andreev reflection occurs in the proximity region of the normal material. Such processes would have been impossible if the surface of Bi-2122 was non-superconducting, or if the contact quality was spoiled by an oxide layer.

So far we have found the method is almost always successful when both samples are properly cleaved to reveal large areas with atomically flat surfaces. In particular, it was found that simply cleaving $Bi_2Se_3$ and $Bi_2Te_3$ with scotch tape only produced high barrier junctions. Whereas when the samples were cleaved with glass slides and tape, large atomically smooth areas were produced and the proximity effects and Andreev reflection were observed. From this we conclude that it is the ability to produce large-area atomically smooth surfaces that enables the production of low-barrier junctions. In particular since the samples are cleaved in inert atmosphere it is likely that they can adhere to one another with no foreign materials in between. Furthermore, it is well known that both materials can be mechanically exfoliated, suggesting that when pressed together these two materials can adhere to one another via Van-der Waals interactions.

A Bi-2212/Graphite mechanically-bonded junction was fabricated by our mechanical bonding method, and AC differential conductance measurements were performed similar to the measurements done on Bi-2212/Bi$_2$Se$_3$ (Supplementary Figure 1). The graphite-based junction tunneling spectrum measurements show the typical high-resistance superconducting gap (Supplementary Figure S1) around 45mV – similar to those measured for Bi-2212/Bi$_2$Se$_3$.